\documentclass{article}

\usepackage{graphicx}
\usepackage{dcolumn}
\usepackage{bm}

\usepackage[utf8]{inputenc}
\usepackage[T1]{fontenc}
\usepackage{mathptmx}

\begin{document}

\title{When is a set of questions to nature together with sharp answers to those questions in one-to-one correspondence with a set of quantum states?}

\author{Inge S. Helland \\ Department of Mathematics, University of Oslo \\ P.O.Box 1053 Blindern, N-0316 Oslo, Norway \\ ingeh@math.uio.no}

\date{\today}

\maketitle

\begin{abstract}

A general question is posed to the quantum community. Partial results are formulated in a self-contained way. In particular, the title question is answered affirmatorily in two cases: 1) The case of spin/ angular momentum of a partcle; 2) A general symmetry situation under certain technical assumptions.

\end{abstract}

Gerhard 't Hooft: Quantum mechanics is the answer. What is the question?

\section{Introduction}

It is now a universal agreement among physicists that quantum mechanics is the most successful physical theory that has ever been developed. The calculations devised from the theory may be complicated, but there is again a universal agreement on how the calculations should be carried out. 

Yet there is no agreement at all when it comes to interpretation of the theory. Many conferences on quantum foundation have been arranged in recent years, but as a result of this, the number of new interpretations has increased, and no one of the old has died out. In two of these conferences, a poll among the participants was carried out [1, 2]. The result was an astonishing disagreement on several simple and fundamental questions. One of these questions was whether quantum theory should be interpreted as an objective theory of the world (the ontic interpretation) or if it only expresses our knowledge of the world (the epistemic interpretation).

The present note takes as a point of departure an epistemic interpretation of quantum theory, as does the book [3]. Chapter 4 in that book contains an error. (The statement on page 63 that Definition 4.8 makes sense in the spin/angular momentum space is wrong; a corresponding error in the paper [4] was pointed out by Yoh Tanimoto.) Nevertheless, the problem taken up in that chapter is of importance, in my opinion. The purpose of this note is to state this problem in a self-contained way.

In general, conceptual variables, variables defined by an observer or by a group of communicating observers, are important. These variables are imaged to emerge in connection to an epistemic process. Some such variables can not be measured, are inaccessible, say the vector (position, momentum) of a particle by Heisenberg's inequality. Those which can be measured, are called epistemic conceptual variable or e-variables. As will be discussed below, the states of the physical system have interpretations related to a question `What is the value of $\theta$?' for an e-variable $\theta$, together with some information about $\theta$, in the simplest, discrete, case full information $\theta=u$. Our main problem is to find out how general this interpretation is.

\section{Interpretation of the quantum state?}

\subsection{Conceptual variables} 

A conceptual variable is any variable defined by an observer or by a group of communicating observers. I will assume that each conceptual variable varies on some topological space. For a physical system under measurement, two kinds of conceptual variables exist. 

A variable may be accessible, possible to measure, like a velocity or a spin-component of a particle in some given direction. Such variables are called epistemic conceptual variables, \emph{e-variables}, and are in my view closely related to the parameters of statistical theory. 

Or they may be inaccessible, like the full spin-vector or the vector (position, velocity). When a vector $\phi=(\theta^1,\theta^2 )$ is inaccessible, but the components $\theta^1$ and $\theta^2$ are e-variables, we are in a situation where we have a choice of measurement, and we might say that $\theta^1$ and $\theta^2$ are complementary.

\subsection{State vectors corresponding to maximally accessible e-variables}

Assume a finite-dimensional Hilbert space $H$, and let $|v\rangle$ be some unit vector in $H$. Then trivially, $|v\rangle$ is an eigenvector of many operators. Assume that one can find such an operator $A$ satisfying 1) $A$ is physically meaningful, that is, can be associated with an e-variable $\theta$; 2) $A$ has only one-dimensional eigenspaces. 

Then in particular, $|v\rangle$ corresponds to a single eigenvalue $u$ of $A$, and can be associated with a question: `What is the value of $\theta$?' together with an answer: `$\theta=u$'.

It is easy to see, and is shown explicitly in [4], that all eigenspaces of $A$ are one-dimensional if and only if $\theta$ is maximally accessible: Whenever $\theta=f(\zeta)$ for a function $f$ which is not one-to-one, the conceptual variable $\zeta$ is inaccessible.

\subsection{General state vectors}

If the situation is as in Subsection 2.2, but $A$ is more general, then the \emph{eigenspaces} of $A$ can be associated with a question-and-answer pair concerning $\theta$.

To see this, let $\lambda$ be a maximally accessible e-variable, and let $\theta=t(\lambda)$ for some function $t$. Let $B$ be the operator corresponding to $\lambda$, and let $B$ have the eigenvalue decomposition

\[B=\sum v_j |j\rangle \langle j|.\]

Then $\{v_j\}$ are the possible values of $\lambda$, and these can be connected to unique question-and-answer pairs. Let now $\{u_i\}$ be the possible values of $\theta=t(\lambda)$. Define $C_i = \{j: t(v_j)=u_i\}$, let $V_i$ be the space spanned by $\{|j\rangle : j\in C_i\}$, and let $\Pi_i$ be the projection upon $V_i$. Then we can define the operator corresponding to $\theta$:

\[A=\sum_i u_i \Pi_i = \sum_i \sum_{j\in C_i } t(v_j) |j\rangle \langle j|.\]

The vector space $V_i$ (or any unit vector in that space) can be interpreted as 1) The question `What is the value of $\theta$?' has been posed. 2) We have obtained the answer `$\theta=u_i$'.

\subsection{Spin and angular momentum}

The interpretation given above is general, but constrained to single e-variables. For the case of spin/angular momentum, more concrete general interpretations can be given, at least for certain state vectors. To see this, consider a spin or angular momentum vector $\phi$ with fixed norm varying on a sphere $\Phi$. This vector is inaccessible. However, given some direction $a$, the components $\theta^a =a\cdot \phi$ can be measured and are e-variables. Given a certain normalization of $\phi$, each $\theta^a$ takes the values $-j, -j+1,...,j-1, j$ for some integer or half-integer $j$.
\bigskip

\textbf{Proposition 1}

\textit{Assume the usual Hilbert space for spin/angular momentum. For each $a$ and each $h$ (h=-j,...,j) there is exactly one normalized ket vector $|v\rangle=|a;h\rangle$ with arbitrary phase such that the operator $J^a$ corresponding to $\theta^a$ satisfies $J^a |v\rangle=h|v\rangle$. This ket vector corresponds to the question `What is the value of the angular momentum component $\theta^a$?' together with the definite answer `$\theta^a =h$'.}
\bigskip

For the qubit case, dimension 2 of the Hilbert space, these vectors constitute all ket vectors. This can be seen by a simple Bloch sphere argument. The proof of this and the proof of Proposition 1 is given in Appendix 1. 

The qubit case has recently been extensitively generalized by H\"{o}hn [5-7]. In those papers the quantum formalism for systems of qubits is reconstructed from elementary rules on an observer's information acquisition: Essentially sets of questions and sharp answers to those questions.

\subsection{A tentative general theorem}

Let in general $\phi$ be an inaccessible conceptual variable taking values in some topological space $\Phi$, and let $\theta^a =\theta^a (\phi)$ be accessible functions for $a$ belonging to some index set $\mathcal{A}$. Assume that each $\theta^a$ is maximally accessible, and assume that there is a one-to-one relationship between the different e-variables: For $a\ne b$ there exists an invertible transformation $k_{ab}$ such that $\theta^b (\phi)=\theta^a (k_{ab}\phi)$ (no summation convention). The spin/angular momentum situation is a special case of this. In general, $\theta^a$ varies over a space $\Theta^a$,

For each $a$, let $G^a$ be the group of automorphisms on $\Theta^a$, and for $g^a \in G^a$ let $k^a$ be any transformation on $\Phi$ (if it exists) for which $g^a \theta^a(\phi)=\theta^a (k^a \phi)$. Contrary to what was claimed on p. 63 of [3], the problem in the angular momentum case is that such transformations do not exist in general in that case. In the following I will assume that transformations $k^a$ exist and are unique for each $a$ and each $g^a$. Then for fixed $a$ they form a group $K^a$.

Let $K$ be the group on $\Phi$ generated by the $K^a$'s and the elements $k_{ab}$.

Make the following assumptions:

1) The group $K$ is a locally compact topological group satisfying weak assumptions such that an invariant measure $\mu$ on $\Phi$ exists.

2) The group generated by products of elements in $K^a ,K^b,....; a,b,...\in\mathcal{A}$ is equal to $K$.

Consider the case where each $\theta^a$ takes a finite $d$ number of values $\{u_i\}$.

Now fix one index $0\in\mathcal{A}$ and consider the Hilbert space

\begin{equation}
H=\{f\in \mathrm{L^2}(\Phi,\mu ): f(\phi)=\tilde{f}(\theta^0 (\phi))\ \mathrm{for\ some}\ \tilde{f}\}.
\label{H}
\end{equation}
This Hilbert space is $d$-dimensional.
\bigskip

\textbf{Theorem 1} 

\textit{Under some extra technical conditions the following holds: For every $a, u_i$ and associated with every indicator function $I(\theta^a (\phi)=u_i )$ there is a vector $|a;i\rangle \in H$. The mapping $I(\theta^a (\phi)=u_i )\mapsto |a;i\rangle$ is invertible in the sense that $|a;i\rangle \ne |b;j\rangle$ for all $a,b,i,j$ except in the trivial case $a=b; i=j$.  This inequality is interpreted to mean that there is no phase factor $\gamma$ such that $|a;i\rangle =\gamma|b;j\rangle$. For each $a$ the vectors $|a;i\rangle$ form an orthonormal basis of $H$.}
\bigskip

This means that the vector $|a;i\rangle$ can be interpreted as a question: `What is the value of $\theta^a$?' together with the answer: `$\theta^a =u_i$'.

The proof of Theorem 1 after specifying one possible set of technical conditions, is given in Appendix 2. This set of conditions do not hold, in fact they are irrelevant, for the spin/ angular momentum case. Nevertheless it is proved in Appendix 1 that Theorem 1 holds in this case (with a different Hilbert space, but all Hilbert spaces of the same dimension are isomorphic). Therefore the above set of conditions can not be the most general one. The open problem raised in this note is to find the weakest possible set of conditions under which Theorem 1 is valid.

\section{Concluding remarks}

The notion of conceptual variables also has links to other interpretations of quantum theory. Take for instance the classical Bohm interpretation, constructed from a particle trajectory plus a pilot wave. These constructions are just conceptual variables, but at least the full trajectory must be inaccessible. Or take the many worlds/ many minds interpretations: Here the different worlds must be considered as conceptual variables, but only one world is accessible.

If a final version of Theorem 1 could be found, this would provide a nice and simple interpretation of at least some states of quantum theory. By the simple observation in subsections 2.2 and 2.3 it should be straightforward to generalize this to all state vectors. The corresponding foundation question is addressed in [4].

 \section*{Acknowledgment}
 
 I am grateful to Bj\o rn Solheim for general discussions.

\section*{References}

\setlength\parindent{0cm}

[1] M. Schlossbauer, J. Koller  and A. Zellinger,  A snapshot of fundamental attitudes toward quantum 
         mechanics. Studies in History and Philosophy of Science \textbf{44}(3), 222-230 (2013)
         
[2] T. Norsen and S. Nelson, Yet another snapshot of fundamental attitudes toward quantum 
          mechanics. arXiv: 1306.4646 [quant-ph] (2013)
          
[3] I.S. Helland, \textit{Epistemic Processes. A Basis for Statistics and Quantum Theory.} (Springer, Berlin, 2018)

[4] I.S. Helland, Symmetry in a space of conceptual variables. J. Math. Phys. 60 (5) 052101 (2019)

[5] P.A. H\"{o}hn, Quantum theory from rules on information acquisition. arXiv: 1612.06849v4 [quant-ph] (2017)

[6] P.A. H\"{o}hn, Toolbox for reconstructing quantum theory from rules on information acquisition. arXiv: 1412.8323v9 [quant-ph] (2018)

[7] P.A. H\"{o}hn and C.S.P. Wever, Quantum theory from questions. arXiv: 1511.01130v7 [quant-ph] (2018)

[8] A. Messiah, \textit{Quantum Mechanics}, Vol. 2. (North-Holland, Amsterdam, 1969)

[9] Z.-Q. Ma, \textit{Group Theory for Physicists.} (World Scientific, Singapore, 2007)

\newpage

\section*{Appendix 1: Proofs in the spin/ angular momentum case.}

Since all separable Hilbert spaces of a given dimension are isomorphic, we are free to and will in this section use the ordinary Hilbert space formulation used in textbooks for angular momenta/spins. The discussion here will rely on Massiah [8]. Let the angular momentum operator be $\mathbf{J}$, let $j$ in $\|\mathbf{J}\|^2=j(j+1)$ be fixed, let $J_x, J_y$ and $J_z$ be the angular momentum operators in directions $x, y$ and $z$, respectively, and let $J_a =a_1 J_x+a_2 J_y+a_3 J_z$ be the operator corresponding to angular momentum $\theta^a$ in the direction $a=(a_1,a_2,a_3)$. Without loss of generality, assume $\sum_i a_i^2=1$. 

\bigskip

\underline{Proof of Proposition 1.} 

Let $\{|m\rangle \}; \ m=-j,...,j$ be the normalized eigenstates of $J_z$, and seek a ket vector $|v\rangle=\sum_{m=-j}^j b_m|m\rangle$ with $\sum |b_m|^2=1$ satisfying $J_a |v\rangle =h|v\rangle$. From [8] the operators $J^+=J_x+iJ_y$ and $J^-=J_x-iJ_y$ satisfy
\[J^+|m\rangle=\sqrt{(j-m)(j+m-1)}|m+1\rangle=A_m|m+1\rangle;\ m\ne j,\]
\[J^-|m\rangle=\sqrt{(j+m)(j-m+1)}|m-1\rangle=B_m|m-1\rangle;\ m\ne -j.\]
Solving this for $J_x$ and $J_y$ leads to
\[J_a|v\rangle=(a_1J_x+a_2J_y+a_3J_z)\sum_{m=-j}^j b_m|m\rangle\]
\[=\sum_{m=-j}^j [\frac{1}{2}(a_1-ia_2)b_{m-1}A_{m-1}+\frac{1}{2}(a_1 +ia_2)b_{m+1}B_{m+1}+a_3 b_m m]|m\rangle\]
if we define $A_{-j-1}=B_{j+1}=0$. Putting this equal to $h\sum b_m|m\rangle$ we get the recursion relations
\[\frac{1}{2}(a_1+ia_2)b_{m+1}B_{m+1}=hb_m-a_3 m b_m-\frac{1}{2}(a_1-ia_2)b_{m-1}A_{m-1}\]
for $h,m=-j,...,j$. 

Put $b_{-j}=c$. Then the recursion relation first gives  $b_{-j+1}=\frac{(h+ja_3)c\sqrt{2}}{(a_1+ia_2)\sqrt{j}}$, and the same relation then determines $b_{m+1}$ for $m=-j+1,...j-1$. The relation for $m=j$ gives an eigenvalue equation for $h$, but we already know from $J^a|v\rangle=h|v\rangle$ that this has solutions $h=-j,-j+1,...,j-1,j$. Finally, $|c|$ is determined from $\sum|b_m|^2=1$. Thus, modulo an arbitrary phase factor, we have a unique ket vector $|v\rangle$ determined from the basis vectors $|m\rangle$, $m=-j,...,j$.
\bigskip

\textbf{Corollary 1}

 \textit{Theorem 1 of Subsection 2.5 is valid for the spin/angular momentum case.}
\bigskip

\underline{Proof.} This follows here from the construction above, giving a solution depending in a unique way upon $a=(a_1,a_2,a_3)$ on the unit sphere.
\bigskip

\textbf{Corollary 2}

 \textit{The epistemic ket vectors of Proposition 1 form a set in the  Hilbert space determined by $h$ and by 2 independent real parameters.}
\bigskip

\underline{Proof.} These simple epistemic states may be indexed by $h$ and by $a_1, a_2$ and $a_3=\pm\sqrt{1-a_1^2-a_2^2}$. 

In the spin 1/2 the unit ket vectors in the spin 1/2 case are determined by the 2-dimensional Bloch sphere. Hence the following result is intuitive from Corollary 2:
\bigskip

\textbf{Proposition 2}

\textit{ For the spin 1/2 case the vectors of Proposition 1 give all unit vectors in the Hilbert space of dimension 2.}

\bigskip

\underline{Proof.} Let $v_0$ be a fixed 2-dimensional complex unit  vector, e.g., $v_0=(1,0)$, and let $v$ be any complex vector of dimension 2. Then there is a unitary matrix $M$ with determinant 1 such that $v=Mv_0$. It is well known (see, e.g., Ma [9]) that there is a homomorphism of the group $SU(2)$ of unitary matrices with determinant 1 onto the 3-dimensional rotation group. Let $R$ be the image of $M$ under this homomorphism. Fix a fixed direction $a_0$, let $a=Ra_0$, and look at the ket vector $|a;+\rangle$. This gives a mapping from $v$ to $|a;+\rangle$. Changing $a$ into $-a$ in this argument gives a mapping from the complex unit 2-vectors to the ket vectors $|a;-\rangle$. By Proposition 1 each ket vector $|a;h\rangle$ is a complex unit vector. Hence we have established a one-to-one correspondence.
\bigskip

For $j>1/2$ the set of simple epistemic states for angular momenta is not closed under linear combinations. As remarked in Subsection 2.2, however, each pure quantum state can nevertheless be seen as the eigenstate of some operator; the problem is only to associate this operator to a physically meaningful e-variable. For $j>1/2$ it is not enough to look at angular momentum components $\theta^a$ in various directions $a$. Candidates for other physically meaningful e-variables may be, e.g., $\alpha \theta^a+\beta \theta^b$ for $a\ne b$, or more generally any $f(\theta^a,\theta^b,\theta^c,...)$.
\newpage

\section*{Appendix 2, Technical conditions and a proof of Theorem 1.}

Fix $0\in\mathcal{A}$, and let $H$ be the Hilbert space
\[H=\{f\in L^2(\Phi,\mu):\ f(\phi)=\tilde{f}(\theta^0(\phi))\ \mathrm{for\ some\ }\tilde{f}\}.\]
Here $L^2(\Phi,\mu)$ is the set of all complex functions $f$ on $\Phi$ such that $\int_\Phi |f(\phi)|^2 d\mu <\infty$. Two functions $f_1$ and $f_2$ are identified if $\int_\Phi |f_1(\phi)-f_2(\phi)|^2 d\mu =0$. From now on 
I will assume that the $\theta^a$'s are discrete. Then $H$ is separable. If the $\theta^a$'s take $d$ different values, $H$ is $d$-dimensional. Since all separable Hilbert spaces are isomorphic, it is enough to arrive at the quantum formulation on this $H$.
\bigskip

\textbf{Lemma 1} 

\textit{The values $u_i^a$ of $\theta^a$ can always be arranged such that $u_i^a=u_i$ is the same for each $a$ ($i=1, 2, ...$).}
\bigskip

\underline{Proof.} By Assumption 1
\[\{\phi:\theta^b=u_i^b\}=\{\phi:\theta^a(k_{ab}\phi)=u_i^b\}=k_{ba}(\{\phi:\theta^a(\phi)=u_i^b\}).\]
The sets in brackets on the lefthand side here are disjoint with union $\Phi$. But then the sets in brackets on the righthand side are disjoint with union $k_{ab}(\Phi)=\Phi$, and this implies that $\{u_i^b\}$ gives all possible values of $\theta^a$.
\smallskip

Let $|a;i\rangle$ be the ket vector asserted to exist in Theorem 1.
When a ket vector is defined, a corresponding bra vector can be defined. The operator corresponding to $\theta^a$ can be defined as
\[A^a =\sum_i u_i |a;i\rangle\langle a;i|.\]
In the maximal setting this has non-degenerate eigenvalues.
\bigskip

\underline{Proof of Theorem 1 under an extra assumption}

Let $U$ be the left regular representation of $K$ on $L^2(\Phi,\mu)$: $U(k)f(\phi)=f(k^{-1}\phi)$. It is well known that this is a unitary representation. We will seek a corresponding representation of $K$ on the smaller space $H$.

In the following, recall that upper indices as in $k^a$ indicate variables related to a particular $\theta^a$, here a group element of $K^a$. Also recall that $0$ is a fixed index in $\mathcal{A}$. Lower indices as in $k_{ab}$ has to do with the relation between two different $\theta^a$ and $\theta^b$.
\bigskip

\textbf{Proposition 3} 

\textit{a) A (multivalued) representation $V$ of $K$ on the Hilbert space $H$ can always be found.}

\textit{b) There is an extended group $K'$ such that $V$ is a univalued representation of $K'$ on $H$.}

\textit{c) There is a homomorphism $K'\rightarrow K^0$ such that $V(k')=U(k^0)$. If $k'\ne e'$ in $K'$, then $k^0\ne e$ in $K^0$.}
\bigskip

\underline{Proof.} a) For each $a$ and for $k^a\in K^a$ define $V(k^a)=U(k_{0a})U(k^a)U(k_{a0})$. Then $V(k^a)$ is an operator on $H$, since it is equal to $U(k_{0a}k^a k_{a0})$, and $k_{0a}k^a k_{a0}\in K^0 =k_{0a}K^a k_{a0}$. For a product $k^a k^b k^c$ with $k^a\in K^a$, $k^b \in K^b$ and $k^c \in K^c$ we define $V(k^a k^b k^c)=V(k^a )V(k^b)V(k^c)$, and similarly for all elements of $K$ that can be written as a finite product of elements from different subgroups. 

Let now $k$ and $h$ be any two elements in $K$ such that $k$ can be written as a product of elements from $K^a, K^b$ and $K^c$, and similarly $h$ (the proof is 
similar for other cases.) It follows that $V(kh)=V(k)V(h)$ on these elements, since the last factor of $k$ and the first factor of $h$ either must belong to 
the same subgroup or to different subgroups; in both cases the product can be defined by the definition of the previous paragraph. In this way we see that $V$ is 
a representation on the set of finite products, and since these generate $K$ by Assumption 2b) , it is a representation of $K$.

Since different representations of $k$ as a product may give different solutions, we have to include the possibility that $V$ may be multivalued.
\smallskip

b) Assume as in a) that we have a multivalued representation $V$ of $K$. Define a larger group $K'$ as follows: If $k^a k^b k^c =k^d k^e k^f$, say, 
with $k^i \in K^i $ for all $i$, we define $k_1'=k^a k^b k^c $ and $k_2'=k^d k^e k^f$. Let $K'$ be the collection of all such new elements that can be written as a formal product of elements $k^i \in K^i$. 
The product is defined in the natural way, and the inverse by for example $(k^a k^b k^c )^{-1}=(k^c )^{-1}(k^b )^{-1}(k^a)^{-1}$. By Assumption 2) in Subsection 2.5, the group $K'$ 
generated by this construction must be at least as large as $K$. It is clear from the proof of a) that $V$ also is a representation of the larger group $K'$ on $H$, now a one-valued representation.
\smallskip

c)  Consider the case where $k' =k^a k^b k^c$ with $k^i \in K^i$. Then by the proof of a):
\[V(k') = U(k_{0a}) U(k^a)U(k_{a0}) U(k_{0b}) U(k^b)U(k_{b0})U(k_{0c}) U(k^c)U(k_{c0}) \]
\[=U(k_{0a}k^a k_{a0} k_{0b}k^b k_{b0} k_{0c} k^c k_{c0})
=U(k^0),\]
where $k^0 \in K^0$. The group element $k^0$ is unique since the decomposition $k'=k^a k^b k^c$ is unique for $k'\in K'$. The proof is similar for other decompositions. By the construction, the mapping $k'\rightarrow k^0$ is a homomorphism.

Assume now that $k^0 =e$ and $k'\ne e'$. Since $U(k^0)\tilde{f}(\theta^0(\phi))=\tilde{f}(\theta^0((k^0)^{-1}(\phi)))$, it follows from $k^0=e$ that $U(k^0)=I$ on $H$. 
But then from what has been just proved, $V(k')=I$, and since $V$ is a univariate representation, it follows that $k'=e'$, contrary to the assumption.
\bigskip

\textbf{Assumption 3)}

\textit{a) $U$ is an irreducible representation of every cyclic subgroup of the group $K^0$ on $H$ other than the trivial group, and the dimension $d$ of $H$ is larger or equal to 2.}

\textit{b) The representation $V$ of the whole group $K$ is really multivalued on the elements $k_{ab}$.}
\bigskip

Now choose an orthonormal basis for $H$: $f_1,...,f_d$ where $f_i(\phi)=\tilde{f}_i(\theta^0(\phi))$, and where the interpretation of $f_i$ is that $\theta^0=u_i$. Write $|0;i\rangle=f_i(\phi)$.
\bigskip

\textbf{Assumption 3)}

\textit{c) When finding this basis, one can choose $ \tilde{f}_i$ and $\tilde{f}_j$ in such a way that there exists a $\theta_1$ such that $\tilde{f}_i(g^0 \theta_1)\ne \tilde{f}_j(\theta_1)$ for all $g^0 \in G^0$.}
\bigskip

\textbf{Lemma 2} 

\textit{For every $i$ and every $k^0\in K^0$, $k^0\ne e$, we have $U(k^0)f_i\ne f_i$ in the sense that the two functions can not be made equal by multiplying with a phase factor.}
\bigskip

\underline{Proof.}
Let $d\ge 2$. Assume that there exist a phase factor $\gamma$ and $k^0\ne e$ such that $U(k^0)f_i = \gamma f_i$. Then $\sqrt{\gamma}f_i$ span a one-dimensional subspace of $H$ which is invariant under the cyclic group generated by $k^0$, contrary to the assumption of irreducibility.
\bigskip

Introduce now the assumption that the representation $V$ really is multivalued. Let $k_{0a1}'$ and $k_{0a2}'$ be two different elements of the group $K'$, both corresponding to $k_{0a}$ of $K$. Define $k_a'=(k_{0a1}')^{-1}k_{0a2}'$. Then $k_a'\ne e'$ in $K'$. By the homomorphism of Proposition 3c), let $k_a'\rightarrow k_a^0$. Then $k_a^0\ne e$ in $K^0$. Now define
\[|a;i\rangle = \tilde{f}_i(\theta^0(k_a^0\phi))=U((k_a^0)^{-1})|0;i\rangle.\]

\underline{Proof of Theorem 1 under Assumption 3).} By Lemma 2, $|a;k\rangle \ne |0;k\rangle$. Here and below, inequality of state vectors is interpreted to mean that they can not be made equal by introducing a phase factor.

Next let $i\ne j$. I will prove that the basis functions $f_1,...,f_d$ can be chosen so that $|a;i\rangle\ne |0;j\rangle$ for all $a$. To this end, choose $\tilde{f}_i$ and $\tilde{f}_j$ in such a way that there exists an $\theta_0^j$ such that $\tilde{f}_i(g^0 \theta_0^j)\ne\tilde{f}_j(\theta_0^j)$ for all $g^0 \in G^0$. Then for any fixed $k$, $\tilde{f}_{ik}$ defined by $\tilde{f}_{ik}(\theta^0(\phi))=\tilde{f}_i(\theta^0(k\phi))$ is different from $\tilde{f}_j$, and $|a;i\rangle\ne |0;j\rangle$ for all $a$.

The proof that $|a;i\rangle\ne |b;j\rangle$ (except in the trivial case $a=b, i=j$) holds under assumption 3, is a simple extension.

The vectors $|0;i\rangle$ are chosen to be an orthonormal basis for $H$. Since $|a;i\rangle=U|0;i\rangle$ for some unitary $U$, it follows that the vectors $|a;i\rangle$ form an orthonormal basis.
\bigskip

\end{document}